%% file: public.tex
\documentclass[a4paper,11pt]{article}
\pdfoutput=1
\usepackage{jheppub}
\usepackage[T1]{fontenc}
\usepackage{multirow}
\input{acronyms}

\title{Sensitivity of an Early Dark Matter Search using the Electromagnetic Calorimeter as a Target for the Light Dark Matter eXperiment}

\collaboration{LDMX Collaboration}
\collaborationImg{\includegraphics{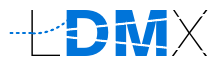}}

\input{author}
\abstract{\input{abstract}}

\begin{document}
\maketitle
\flushbottom

\section{Introduction}
\label{sec:intro}
\input{Introduction}

\section{The LDMX Detector}
\label{sec:ldmx}
\input{ldmx}

\section{Simulated Samples}
\label{sec:sample}
\input{Samples}

\section{Analysis Methodology}
\label{sec:analysis}
\input{Analysis}

\section{Results}
\label{sec:results}
\input{results}

\section{Conclusions}
\label{sec:conclusions}
\input{Conclusion}

\acknowledgments
Support for UCSB is made possible by the Joe and Pat Yzurdiaga endowed chair in experimental science. 
Use was made of the UCSB computational facilities administered by the Center for Scientific Computing at the California NanoSystems Institute and Materials Research Laboratory (an NSF MRSEC; DMR-1720256) and purchased through NSF CNS-1725797, and from  
resources provided by the Swedish National Infrastructure for Computing at the Centre for Scientific and Technical Computing at Lund University (LUNARC), as well as LUNARC’s own infrastructure. Contributions from Caltech, CMU, Stanford, TTU, UMN, UVA, and UCSB are supported by the US Department of Energy under grants DE-SC0011925, DE-SC0010118, DE-SC0022083, DE-SC0015592, DE-SC00012069, DE-SC0007838, and DE-SC0011702, respectively. Support for Lund University is made possible by the Knut and Alice Wallenberg foundation (project grant Light Dark Matter, Dnr. KAW 2019.0080), and by the Crafoord foundation (Dnr 20190875) and the Royal Physiographic Society of Lund.  RP acknowledges support through the L’Or\'{e}al-UNESCO For Women in Science in Sweden Prize with support of the Young Academy of Sweden, and from the Swedish Research Council (Dnr 2019-03436). LB acknowledges support from the Knut and Alice Wallenberg Foundation Postdoctoral Scholarship Program at Stanford (Dnr. KAW 2018.0429). JE, GK, CH, WK, CMS, and NT are supported by the Fermi Research Alliance, LLC under Contract No. DE-AC02-07CH11359 with the U.S. Department of Energy, Office of Science, Office of High Energy Physics and the Fermilab LDRD program.  CB, PB, OM, TN, PS, and NT are supported by Stanford University under Contract No. DE-AC02-76SF00515 with the U.S. Department of Energy, Office of Science, Office of High Energy Physics.

\bibliographystyle{JHEP}
\bibliography{bibliography}

\end{document}

%% file: acronyms.tex
\usepackage{acronym}
\acrodef{ldmx}[LDMX]{Light Dark Matter eXperiment}
\acrodef{ecal}[ECal]{Electromagnetic Calorimeter}
\acrodef{hcal}[HCal]{Hadronic Calorimeter}
\acrodef{eot}[EoT]{electrons-on-target}
\acrodef{bdt}[BDT]{Boosted Decision Tree}
\acrodef{pn}[PN]{photon-nuclear}
\acrodef{en}[EN]{electron-nuclear}
\acrodef{dm}[DM]{dark matter}
\acrodef{pe}[PE]{photo-electron}
\acrodef{db}[DB]{dark bremsstrahlung}
\acrodef{eat}[EaT]{ECal as Target}
\acrodef{rms}[RMS]{root-mean-squared spread}
\acrodef{mm}[MM]{Missing Momentum}

\usepackage{xspace}

\newcommand{\ecal}{ECal\xspace}
\newcommand{\hcal}{HCal\xspace}
\newcommand{\GeV}{\ensuremath{\,\text{Ge\hspace{-.08em}V}}\xspace}
\newcommand{\MeV}{\ensuremath{\,\text{Me\hspace{-.08em}V}}\xspace}

\usepackage{siunitx}
\newcommand{\fourgev}{\qty{4}{GeV}\xspace}
\newcommand{\eightgev}{\qty{8}{GeV}\xspace}

\usepackage{relsize}
\newcommand\babar{%
  \mbox{\slshape B\kern-0.1em{\smaller A}\kern-0.1em B\kern-0.1em{\smaller A\kern-0.2em R}}%
  \xspace
}

%% file: author.tex
\author[a]{Torsten~Åkesson}
\author[b]{Elizabeth~Berzin}
\author[c]{Cameron~Bravo}
\author[d]{Liam~Brennan}
\author[a]{Lene~Kristian~Bryngemark}
\author[c]{Pierfrancesco~Butti}
\author[d]{Filippo~Delzanno}
\author[e]{E.~Craig~Dukes}
\author[f]{Valentina~Dutta}
\author[g]{Bertrand~Echenard}
\author[e]{Ralf~Ehrlich}
\author[h]{Thomas~Eichlersmith\footnote{Corresponding author}}
\author[a]{Einar~El\'{e}n}
\author[h]{Andrew~Furmanski}
\author[g]{Victor~Gomez}
\author[c]{Matt~Graham}
\author[d]{Chiara~Grieco}
\author[e]{Craig~Group}
\author[a]{Hannah~Herde}
\author[i]{Christian~Herwig}
\author[g]{David~G.~Hitlin}
\author[e]{Tyler~Horoho}
\author[d]{Joseph~Incandela}
\author[g]{Nathan~Jay}
\author[d]{Asahi~Jige}
\author[j]{Wesley~Ketchum}
\author[j]{Gordan~Krnjaic}
\author[d]{Amina~Li}
\author[d]{Zihan~Ma}
\author[h]{Jeremiah~Mans}
\author[e]{Cristina~Mantilla~Suarez}
\author[d]{Sanjit~Masanam}
\author[d]{Phillip~Masterson}
\author[h]{Steven~Metallo}
\author[g]{Sophie~Middleton}
\author[h]{Joseph~Muse}
\author[c]{Timothy~Nelson}
\author[b]{Rory~O'Dwyer}
\author[g]{James~Oyang}
\author[e]{Jessica~Pascadlo}
\author[c]{Emrys~Peets}
\author[a]{Luis~Sarmiento~Pico}
\author[a]{Ruth~Pöttgen}
\author[c]{Philip~Schuster}
\author[d]{Chris~Sellgren}
\author[b]{Lauren~Tompkins}
\author[c]{Natalia~Toro}
\author[j]{Nhan~Tran}
\author[d]{Tamas~Vami}
\author[a]{Erik~Wallin}
\author[d]{Yuxuan~Wang}
\author[j]{Andrew~Whitbeck}
\author[d]{Duncan~Wilmot}
\author[d]{Xinyi~Xu}
\author[d]{Danyi~Zhang}

\affiliation[a]{Lund University, Department of Physics, Box 118, 221 00 Lund, Sweden}
\affiliation[b]{Stanford University, Menlo Park, CA 94025, USA}
\affiliation[c]{SLAC National Accelerator Laboratory, Menlo Park, CA 94025, USA}
\affiliation[d]{University of California at Santa Barbara, Santa Barbara, CA 93106, USA}
\affiliation[e]{University of Virginia, Charlottesville, VA 22904, USA}
\affiliation[f]{Carnegie Mellon University, Pittsburgh, PA 15213, USA}
\affiliation[g]{California Institute of Technology, Pasadena, CA 91125, USA}
\affiliation[h]{University of Minnesota, Minneapolis, MN 55455, USA}
\affiliation[i]{University of Michigan, Ann Arbor, MI 48109, USA}
\affiliation[j]{Fermi National Accelerator Laboratory, Batavia, IL 60510, USA}

\emailAdd{eichl008@umn.edu}

%% file: abstract.tex
The Light Dark Matter eXperiment (LDMX) is proposed to employ a thin tungsten target and a multi-GeV
electron beam to carry out a missing momentum search for the production of
dark matter candidate particles.
We study the sensitivity for a complementary missing-energy-based search using the
LDMX Electromagnetic Calorimeter as an active target with a focus on early running.
In this context, we construct an event selection from a limited set of variables that projects sensitivity
into previously-unexplored regions of light dark matter phase space
-- down to an effective dark photon interaction strength $y$ of approximately $2\times10^{-13}$
($5\times10^{-12}$) for a 1\,MeV (10\,MeV) dark matter candidate mass.

%% file: Introduction.tex
The origin and nature of \ac{dm} remains among the most elusive areas of modern physics \cite{new-ideas-in-dm-2017, dark-sectors-2016, BRN-DMNI-2018,RF6_BI1-2022,SnowmassRF6-2022}.
The a priori extremely large viable mass range of DM candidates can be significantly narrowed by
assuming thermal contact of DM with ordinary matter in the early universe \cite{feebly-interacting-2021,fips-workshop-2020}.
The MeV to GeV mass region of thermal \ac{dm} remains relatively unexplored,
corresponds to the range where other matter is stable over the lifetime of the universe,
and can be probed in accelerator-based fixed target experiments \cite{sub-gev-dm-2012,ldm-2015}.

The planned \ac{ldmx} \cite{whitepaper, snowmass23} is designed to observe the production of sub-GeV thermal \ac{dm}. In the nominal scenario, electrons incident on a thin target undergo a \ac{db} process (inset in Figure~\ref{fig:ldmx}) where non-interacting \ac{dm} particles are radiated through a dark mediator. The detector design allows for measuring the missing momentum carried by the \ac{dm} particles through the measurement of the recoil electron track.  It includes a spectrometer dipole; a silicon tracker with sections before and after the thin target; a high-granularity Si-Tungsten \acf{ecal} with a depth of forty radiation lengths; and a \acf{hcal} as well as a scintillator-bar detector for electron counting. \ac{ldmx} will take data with a high-rate, low-intensity electron beam with a typical current of one electron every \qty{27}{\ns} extracted from the Linac Coherent Light Source II (LCLS-II) at SLAC \cite{lcls-ii}. The primary beam energy is expected to be \eightgev using a planned upgrade to LCLS-II, however some data may be acquired with \fourgev depending on the accelerator and detector schedules.

The use of a thin target requires $4\times 10^{14}$ \ac{eot} to obtain the sensitivity projected for the first phase of \ac{ldmx} \cite{pnpaper}.
However, the majority of beam electrons pass through the target without significant interactions and then enter the \ac{ecal} at full energy. Given the density of material composing the \ac{ecal} and the fine-grained \ac{ecal} readout, the \ac{ecal} can be considered as a second effective target mass for \ac{ldmx}.  In this approach, the components of the recoil electron momentum cannot be individually measured, which results in a measurement of missing energy instead of missing momentum.
Given the thickness of the \ac{ecal}, the effective luminosity of such a measurement can be higher than that with the thin target, though the available handles for managing background are fewer.  This paper quantifies the sensitivity of a search using the \ac{ldmx} \ac{ecal} as a target for a first result with $10^{13}$ \ac{eot}, equivalent to about two weeks of nominal \ac{ldmx} beam exposure.
Scenarios where this initial exposure results from both a \fourgev or \eightgev beam are discussed in this paper.

The paper is organized as follows: in Section II the \ac{ldmx} detector design used in this study is reviewed, as well as the signal and background characteristics. The samples of simulated events are summarized in Section III. In Section IV, the analysis strategy is described. The results and conclusions are discussed in Section V and VI.

%% file: ldmx.tex
\ac{ldmx} will utilize a high-rate, low-current electron beam incident on a thin, solid (tungsten) target to search for light thermal \ac{dm}. The behavior of \ac{dm} production is benchmarked by a dark bremsstrahlung process, where electrons emit a dark mediator or dark photon ($A'$) suppressed by the mixing strength~$\epsilon \ll 1$. In turn, the $A'$ subsequently decays to non-interacting \ac{dm} particles ($\chi$) which escape the detector leaving an energy imbalance relative to the initial beam energy.

\begin{figure}[htb]
    \begin{center}
        \includegraphics[width=\textwidth]{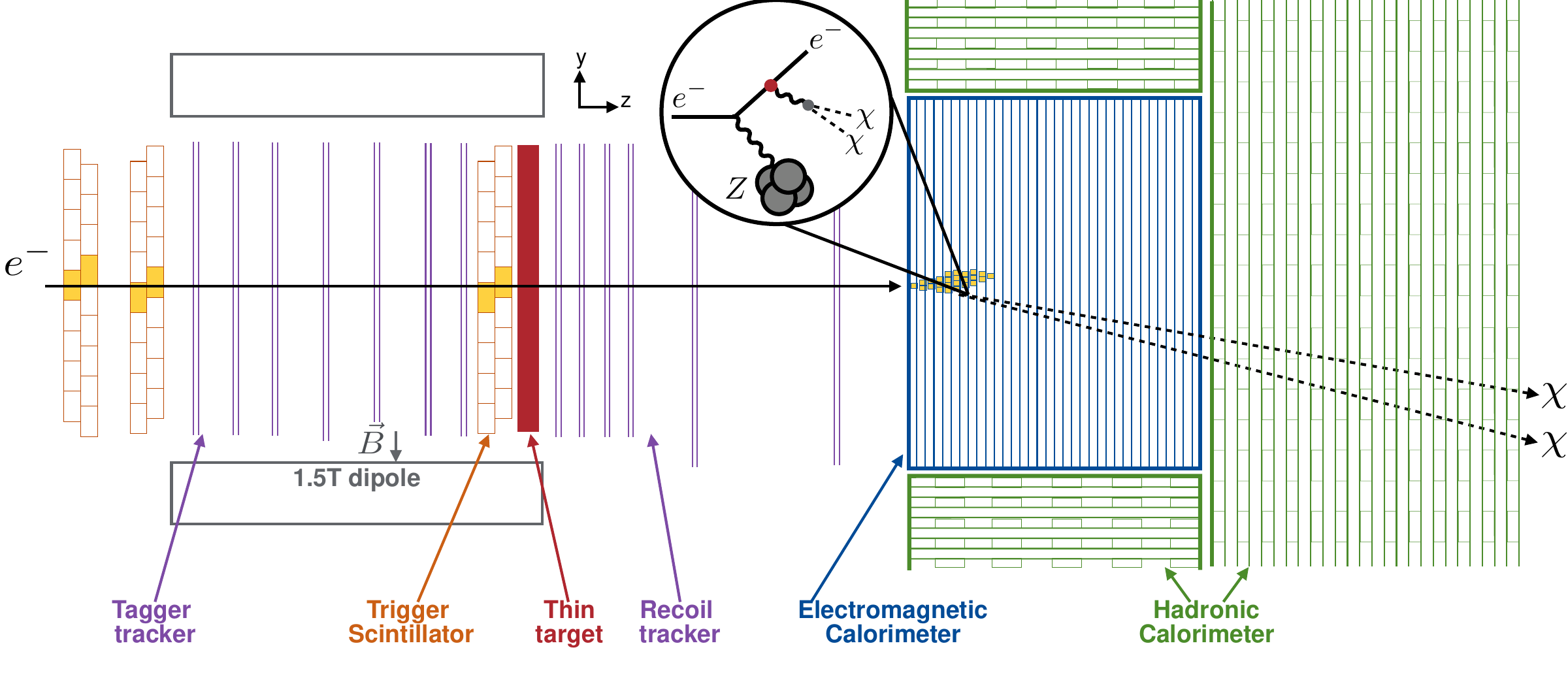}
    \end{center}
    
    \caption{A diagram of the \ac{ldmx} detector apparatus, illustrating production of DM in the \ac{ecal} from a scattering electron, and the corresponding response of the various sub-systems.}
    \label{fig:ldmx}
\end{figure}

The \ac{ldmx} detector, as diagrammed in Figure~\ref{fig:ldmx}, is designed to reject the high rate of background events, including rare \ac{pn} and \ac{en} reactions.
The tracking system consists of a tagging tracker utilizing a \qty{1.5}{T} magnetic field to measure the electron's incident momentum and a recoil tracker utilizing the magnet's fringe field to measure the momentum of charged particles leaving the target.
The \ac{ecal} measures the recoil electron energy as well as interactions that may result from other forward particles that are produced in background processes.
In addition, a sampling \ac{hcal} composed of steel absorber layers and scintillator bars read out by silicon photo-multipliers provides additional background rejection capabilities particularly for events with energetic neutral hadrons.

\begin{figure}
    \begin{center}
        \includegraphics[width=0.75\textwidth]{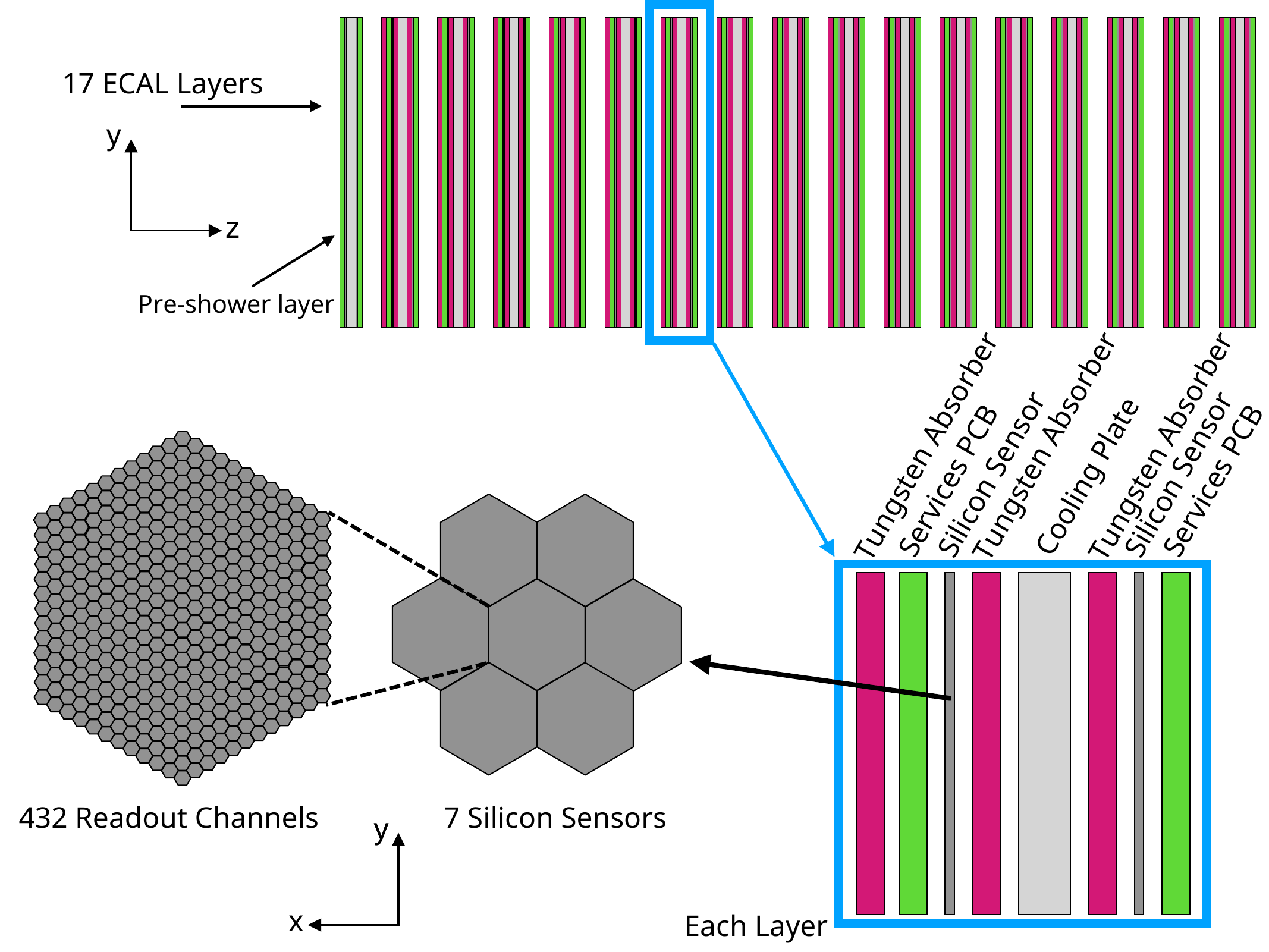}
    \end{center}
    \caption{A diagram of the \ac{ldmx} \ac{ecal} system.
    The longitudinal layer structure is shown (top)
    along with an exploded view of a single double-layer (bottom-right).
    The transverse view of the flower petal arrangement of modules
    as seen from the direction of the beam (bottom-left).}
    \label{fig:ecal}
\end{figure}

The \ac{ecal} design for \ac{ldmx} is adopted from the CMS High Granularity Calorimeter for the CMS Phase-2 upgrade \cite{cms-phase-2-tdr} which fulfills requirements for a fast, efficient, and finely-segmented electromagnetic calorimeter system. The \ac{ecal}, diagrammed in Figure~\ref{fig:ecal}, is a sampling calorimeter with 34 layers organized into 17 double-sided planes, corresponding to approximately 40 $X_{0}$ of material. This large material budget contributes to both an increased rate of \ac{db} production relative to the 0.1 $X_{0}$ thin target, as well as excellent electromagnetic shower containment. Each \ac{ecal} layer contains tungsten absorbers as well as printed circuit boards (PCB) for services and data transport to and from silicon sensors. The sensors each contain 432 readout pads and there are seven sensors arranged in a ``flower'' (Figure \ref{fig:ecal}), providing precise transverse shower reconstruction capable of distinguishing individual electromagnetic showers with small angular separation.  The first pre-shower layer is not preceded by tungsten absorber in order to prevent the shower from beginning before a sensitive layer.

The primary physics trigger of \ac{ldmx} is based on the observation of missing energy in the \ac{ecal}.
For a single incoming electron, the trigger requires the total reconstructed energy in the first twenty layers of the \ac{ecal} ($E_{20}$) to be no more than 1.5 (3.16)~GeV for the 4 (8) GeV beam.
Multi-electron events can be removed from the sample by checking the Trigger Scintillator, Tagger Tracker, and Recoil Tracker subsystems for multiple tracks.
These trigger thresholds were optimized for the missing momentum analysis search channel
in both the \fourgev \cite{pnpaper} and \eightgev \cite{pn8gev} beam cases
and found appropriate for this \ac{eat} missing energy search as well.

%% file: Samples.tex
Samples of events corresponding to several physical processes are simulated in order to
design the missing energy analysis and estimate the sensitivity of this technique to DM production.
The samples used for this study were generated with version 10.2.3 of the \textsc{Geant}4 detector simulation toolkit \cite{geant} with patches to nuclear interaction cross sections and kinematics that are important for \ac{ldmx}'s beam energies \cite{whitepaper} utilizing LDCS \cite{ldcs_2021} for large-scale sample production.
The background and signal samples share the same beam assumptions, namely that electrons arrive upstream of the tagging tracker system with a kinetic energy of 4 or \eightgev.
Each event is simulated with exactly one incoming beam electron.

In order to avoid simulating events that would be removed with a 
tracker-based requirement for the electron to have a certain minimum momentum,
events where the primary electron falls below $87.5\%$ of the beam energy before arrival at the face of the \ac{ecal}
are rejected in all samples.
Based on simulation studies, we expect 62\% of electrons which enter \ac{ldmx} at full energy
to meet this requirement.

Dedicated samples are produced for the signal and unbiased background processes, which are further
complemented by dedicated samples for showers that impart significant momentum into nuclei systems (``enriched nuclear'') or produce muons (``di-muon'').
More efficient production of simulation events is achieved by artificially increasing the cross section of chosen processes by some factor (the biasing factor) for particles above a certain energy threshold (the biasing threshold).
For example, in the photo-nuclear background simulation, the photo-nuclear interaction cross section is increased
for photons whose energy is above the biasing threshold.
The event generation setup is also configured to simulate the interactions of all particles with total energy above the biasing threshold before any other particles in order to improve computational efficiency. Table~\ref{tab:sample_reference} outlines the biasing and filtering for each of the samples used in this study and are explained in more detail in the following sections. Each of the samples was generated and analyzed separately. The unbiased background contains nuclear and di-muon background events, which are rare in the sample but can still be used to validate the biasing and filtering infrastructure. Since the nuclear and di-muon backgrounds are the dominant backgrounds for this analysis channel, the analysis methods are developed mainly using the biased samples for those background processes.

\begin{table}[htb]
    \begin{center}
    \input{samples}
    \end{center}
    \caption{Configuration of the simulation samples used in this analysis. $E_\mathrm{beam}$ is the beam energy being studied (4 or \eightgev in this work). $m_A$ is the mass of the A' in MeV, $\epsilon$ is the dark photon mixing strength, $E_e$ is the energy of the primary electron at the front of the ECal, $E_{A'}$ is the energy of the generated A', $E_{\text{nuc}}$ is the total energy transferred to nuclear interactions during the event, and $E_{\mu}$ is the total energy of produced muons.}
    \label{tab:sample_reference}
\end{table}

\subsection{Signal}

The signal samples were generated using the G4DarkBreM \cite{Eichlersmith:2022bit} technique
in order to account for electron energy loss prior to undergoing a dark bremsstrahlung.
The produced dark photon masses were 1, 5, 10, 50, 100, 500, and 1000 MeV.
The production rate for masses above \qty{100}{\MeV} drops substantially
necessitating an increased biasing factor as a function of mass (Table~\ref{tab:sample_reference})
and leading to weaker sensitivities.
When comparing these samples to other \ac{dm} search literature,
it is common to parameterize the \ac{dm} phase space with an
effective interaction strength $y$ and the mass of the candidate \ac{dm}
$m_\chi$ that would constitute the astronomical \ac{dm} observed today.
Here, these parameters relate to the \ac{dm} model used within the simulation as
$$
  y = \alpha_D \epsilon^2 \left(\frac{m_\chi}{m_{A'}}\right)^4
$$
where $\alpha_D$ is the coupling between $A'$ and $\chi$ and we make standard,
benchmark choices $\alpha_D = 0.5$ and $m_{A'} = 3m_\chi$ connecting
the free parameters $(y,m_\chi)$ and $(\epsilon,m_{A'})$.

In addition to the selection on the primary electron's energy upon entering the \ac{ecal}, events were also required to contain a dark photon with an energy of at least $50\%$ of the beam energy produced in the \ac{ecal} to avoid phase space outside of the trigger acceptance and improve computational efficiency.

\begin{figure}
    \begin{center}
         \includegraphics[width=0.48\textwidth]{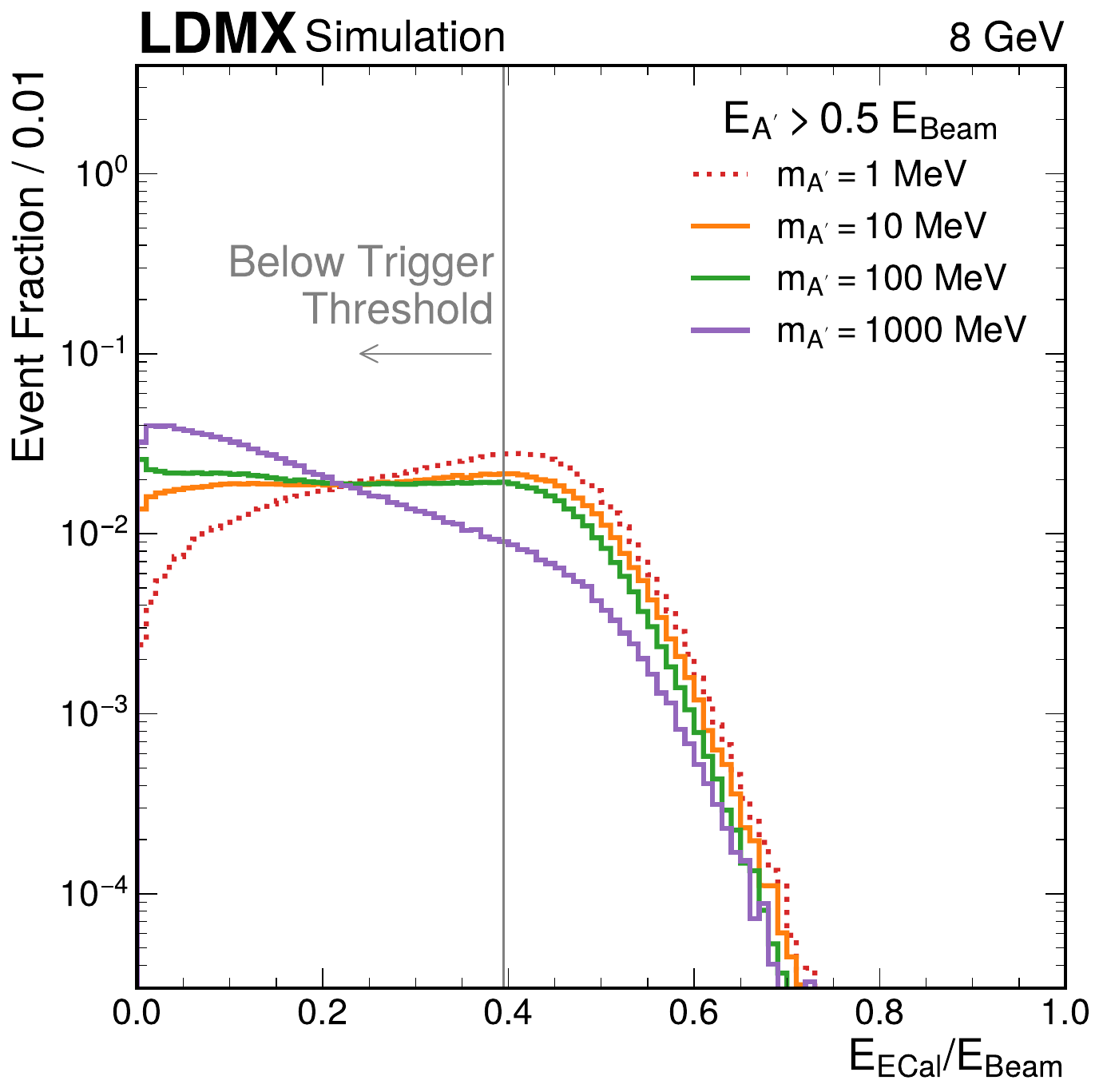}
    \end{center}
    \caption{The total reconstructed energy summed over all layers in the \ac{ecal} ($E_\text{ECal}$) as a fraction of the beam energy ($E_\text{Beam}$) for selected mass points,
    shown for the \eightgev beam case. 
    Heavier dark photons carry a larger amount of energy from the incident electron resulting in a shift of the total reconstructed energy in the \ac{ecal} towards smaller values.
    The gray line indicates the trigger threshold.
    All distributions are normalized so their integral equals one.}
    \label{fig:signal}
\end{figure}

Figure~\ref{fig:signal} shows the total energy deposited in
the \ac{ecal} as a fraction of the beam energy without trigger requirements for selected mass points
for the \eightgev beam case, with the \fourgev beam case being very
similar. The amount of energy deposited in the \ac{ecal} for dark
bremsstrahlung events is dependent on the mass of the dark photon.
Dark photons with larger masses carry a larger amount of energy from the
incident electron resulting in a shift of the total energy reconstructed in
the \ac{ecal} towards smaller values.

\subsection{Enriched Nuclear Background}

While the majority of interactions in the electron shower are electromagnetic, producing photons, electrons, and positrons, hadronic showers are occasionally produced from \ac{en} and \ac{pn} interactions with nuclei in the \ac{ecal}.
The response of the \ac{ecal} to these particles can be significantly smaller than for pure electromagnetic showers.
Hadrons have significantly longer interaction distances in the dense materials of \ac{ecal}
compared to photons, electrons, and positrons, which can result in them escaping the calorimeter.
For this reason, both \ac{en} and \ac{pn} interactions are prominent backgrounds for this analysis.
Since both electrons and photons exist within the developing shower, one cannot easily separate these interactions into distinct processes in the \ac{eat} search; therefore, both types of nuclear interactions are included in this ``enriched nuclear" background.

For the simulation, both \ac{en} and \ac{pn} interactions are biased by the same factor of $200$ in order to maintain relative differences in rates. The energy threshold above which particles are biased is set to $37.5\%$ of the beam energy. A running total of the energy transferred from the initiating electron or photon to the nucleus (so-called ``nuclear energy") is kept in each step of the simulation. 

\begin{figure}[htb]
    \begin{center}
        \includegraphics[width=0.48\textwidth]{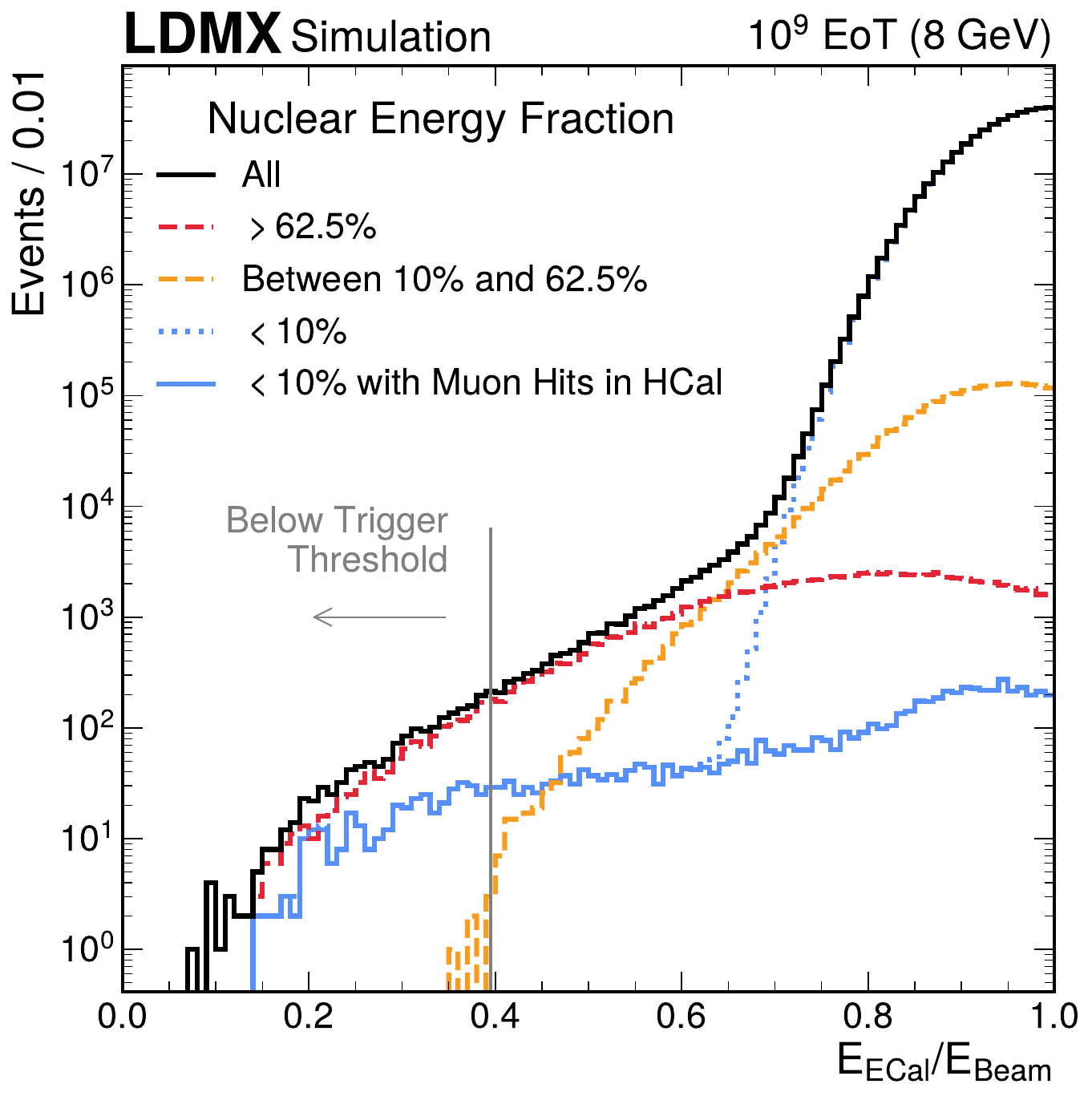}
    \end{center}
    \caption{Total reconstructed energy in the ECal ($E_\text{ECal}$) as a fraction of the beam energy ($E_\text{Beam}$) depending on the total amount of simulated energy transferred to nuclear interactions (the ``Nuclear Energy Fraction'') for an unbiased sample of $10^9$ \ac{eot}. 
    The maximum ECal energy allowed to pass the trigger is drawn in light gray.
    The elongated tail of the blue distribution (Nuclear Energy Fraction $< 10\%$) is
    due to photon conversion to muon pairs as evidenced by simulated energy deposits caused by muons within
    the \ac{hcal} (``Muon Hits in \ac{hcal}'').
    Events with $E_\text{ECal} > E_\text{Beam}$ due to resolution effects are omitted from this plot.}
    \label{fig:nuclear_slices}
\end{figure}

Only events where the total nuclear energy for all particles above the biasing threshold is greater than $62.5\%$ of the beam energy are kept. The $62.5\%$ energy threshold was chosen based on the features displayed in Figure \ref{fig:nuclear_slices} which shows distributions of the total \ac{ecal} energy based on the amount of nuclear energy in unbiased background events.

In order to avoid confining processes to a single nuclear interaction, the biasing and sorting threshold needs to be much less than the $62.5\%$ total nuclear energy threshold.
Moreover, the biasing and sorting threshold needs to be high enough to avoid simulating the entire shower before determining if an event should be kept.
The value of $37.5\%$ was found to be a good balance between these physical and computational bounds.

Figure~\ref{fig:nuclear_slices} also highlights another background process for this analysis: photo-production of muon pairs.
All of the simulated events in the low-nuclear-energy distribution below the trigger threshold are due to photon conversion to two muons.

\subsection{Di-Muon Background}

Simulating the di-muon background is simpler than the enriched nuclear background since it is a single process. The process is biased by a factor of $10^5$ ($10^4$) for \fourgev (\eightgev) beam samples and only photons above half the beam energy are biased. The total energy of both muons produced ($E_{\mu}$) must be greater than half the beam energy in order for the event to be kept.

%% file: samples.tex
\begin{tabular}{|c|c|c|}
    \hline
    Sample & Biasing Factor & Biasing Threshold
    \\ \hline
    Signal & $m_{A'}^{\max(\log_{10}(m_A),2))}/\epsilon^2$ & $0.5E_\text{Beam}$
    \\
    Enriched Nuclear & $200$ & $0.375E_\text{Beam}$
    \\
    Di-Muon & $10^5$ & $0.5E_\text{Beam}$
    \\
    \hline \hline
    Sample & \multicolumn{2}{c|}{Filtering Cuts}
    \\ \hline
    Unbiased  
        & \multicolumn{2}{c|}{$E_e \geq 0.875E_\text{Beam}$}
    \\
    Signal 
        & \multicolumn{2}{c|}{$E_e \geq 0.875E_\text{Beam}$ \& $E_{A'}\geq 0.5E_\text{Beam}$}
    \\
    Enriched Nuclear
        & \multicolumn{2}{c|}{$E_e \geq 0.875E_\text{Beam}$ \& $E_{\text{nuc}}\geq 0.625E_\text{Beam}$}
    \\
    Di-Muon
        & \multicolumn{2}{c|}{$E_e \geq 0.875E_\text{Beam}$ \& $E_{\mu}\geq 0.5E_\text{Beam}$}
    \\ \hline
\end{tabular}

%% file: Analysis.tex
\begin{figure}[htb]
    \begin{center}
      \includegraphics[width=0.48\textwidth]{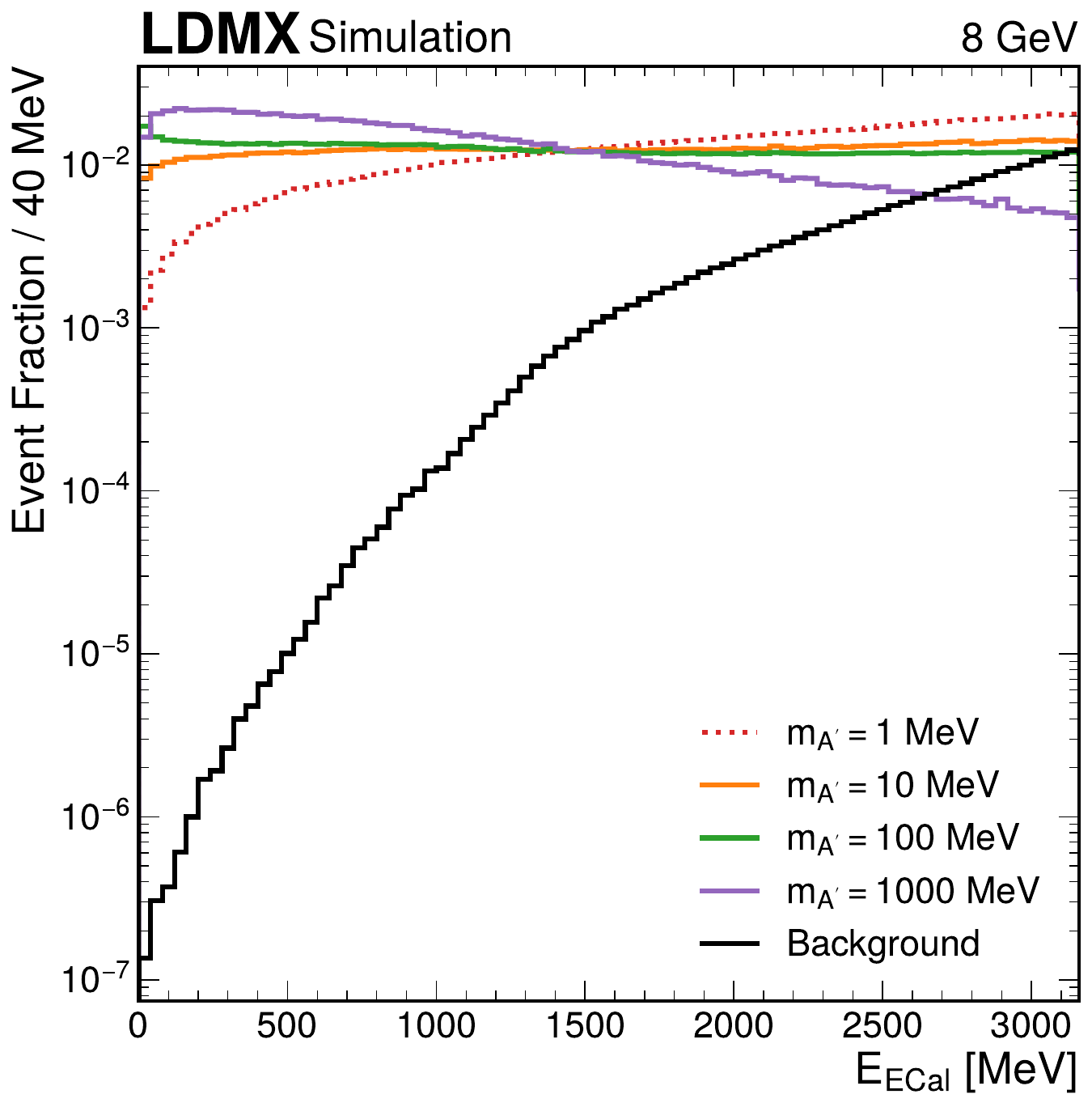}
    \end{center}
    \caption{
    The total reconstructed energy summed over all layers of the ECal ($E_\text{ECal}$)
    for all events that pass the trigger threshold.
    The signal and background distributions are normalized such that their integral is one.
    The events falling into bins with energy above the trigger threshold
    are omitted from this plot but included in efficiency calculations and the normalization.
    }
    \label{fig:ecal_rec_energy}
\end{figure}

The signature of a dark bremsstrahlung event in the electromagnetic
calorimeter is a single electromagnetic shower with significant energy
loss relative to the beam energy.
Figure~\ref{fig:ecal_rec_energy} compares background to various signal
mass points using the unit area normalized distributions of total
reconstructed energy in the \ecal summed over all layers.
As the \ac{eat} search is designed to operate with early data samples when the
trigger energy scale may not be fully calibrated, the energy
requirement for event selection is tightened at the analysis level;
the sum for the energy is performed over all 34 layers of the calorimeter
and the threshold is set \qty{400}{\MeV} lower than what was required by the trigger.
Therefore, for the \eightgev beam the fully-calibrated sum over all layers $E_\text{ECal}$ must be less than \qty{2.76}{\GeV}, while for the \fourgev beam $E_\text{ECal}$ must be less than \qty{1.1}{\GeV}.
The sample with a full energy sum greater than the
selection threshold but an $E_{20}$ below the trigger threshold will
serve as a side band sample for evaluating the prediction for a range
of backgrounds.

After the total energy requirement, the primary remaining backgrounds are those where a significant fraction of the energy is carried by neutrons, $K_L$, or muons through the \ecal.
To suppress these backgrounds, a requirement is placed on \hcal that no scintillator bar in the detector is observed to contain a readout signal corresponding to more than 10 \acp{pe}: $\mathrm{max(PE_{HCal})}<10$.
For the side-\hcal the measurement is made from a single photo-detector, while for the back-\hcal the requirement is placed on the summed readout signal from the photo-detectors on both ends of the bar.
The typical signal for a through-going muon in a single bar is 80 \ac{pe} summed over both ends
while the typical noise in a bar is 1 \ac{pe}.

Some \ac{pn} reactions produce several low-energy particles that range out in the \ecal,
often distributing a small amount of energy over a range of cells.
To suppress such events, an additional upper limit on the overall width of the energy
deposition in the \ecal is applied.
Each channel in the \ecal which has an energy deposit above the readout threshold
(corresponding to 50\% of the energy deposited by a muon passing through a cell)
is called a ``hit''.
The overall width of the energy deposition is then measured as the energy-weighted \ac{rms}
of the hit locations in the transverse plane, which quantifies the size of the event
transverse to the particle impact direction on the calorimeter.
This variable is typically larger for \ac{pn} and \ac{en} backgrounds than for the single truncated
electromagnetic shower present for a signal event. 
The event \ac{rms} is required to be less than 20~mm.

\begin{figure}
  \begin{center}
    \begin{tabular}{cc}
        \includegraphics[width=0.48\textwidth]{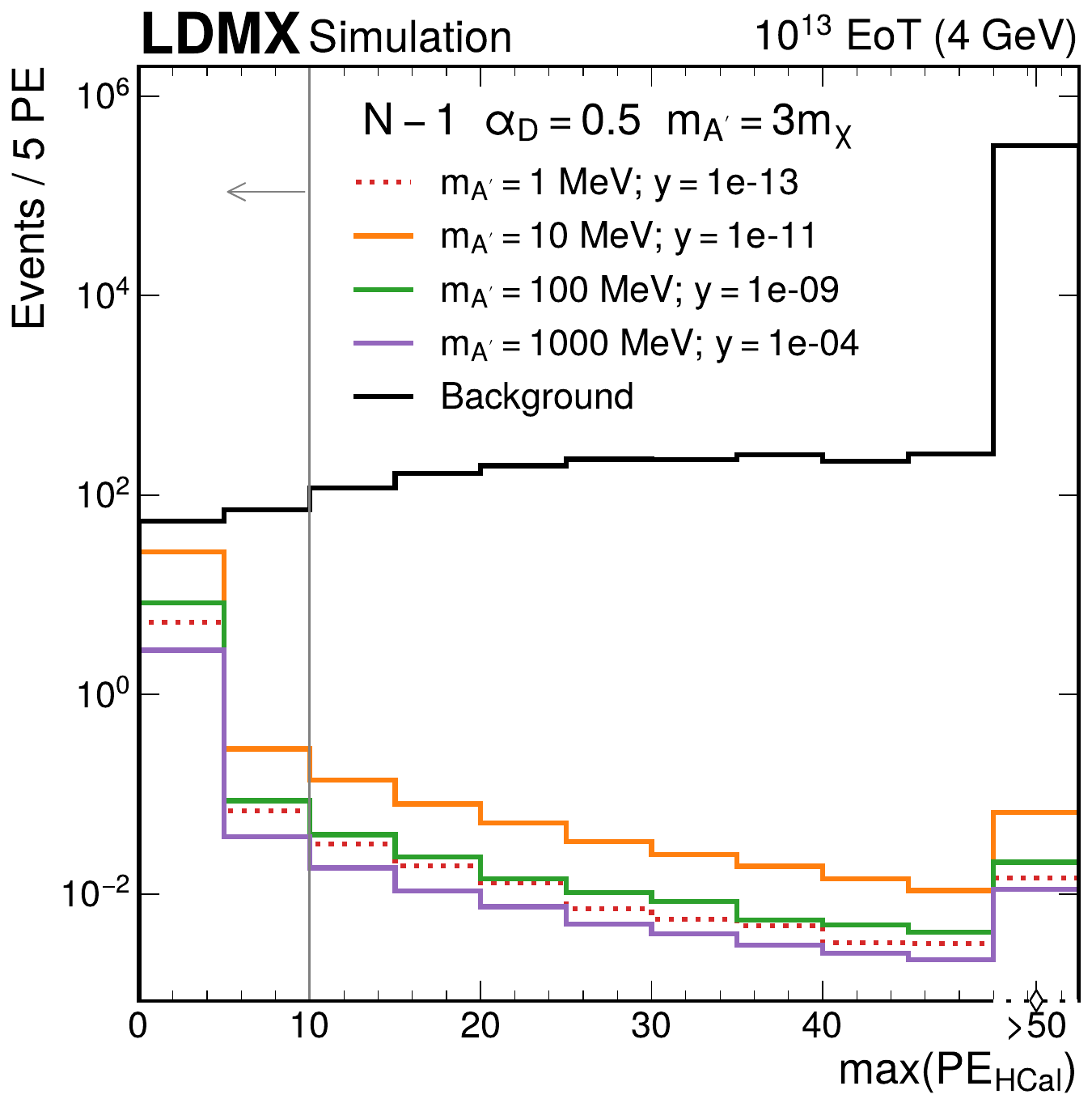} &
        \includegraphics[width=0.48\textwidth]{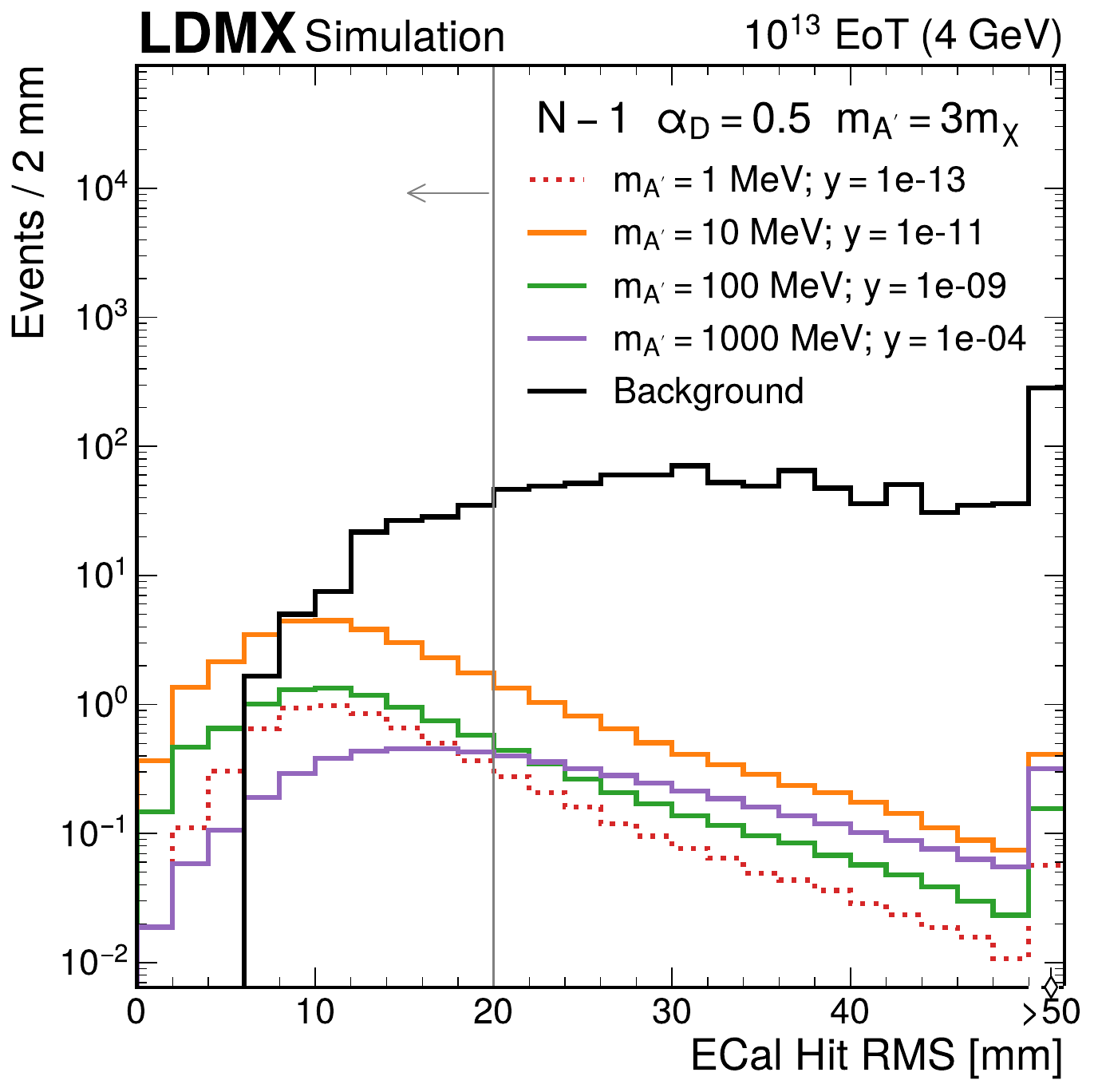} \\
        \includegraphics[width=0.48\textwidth]{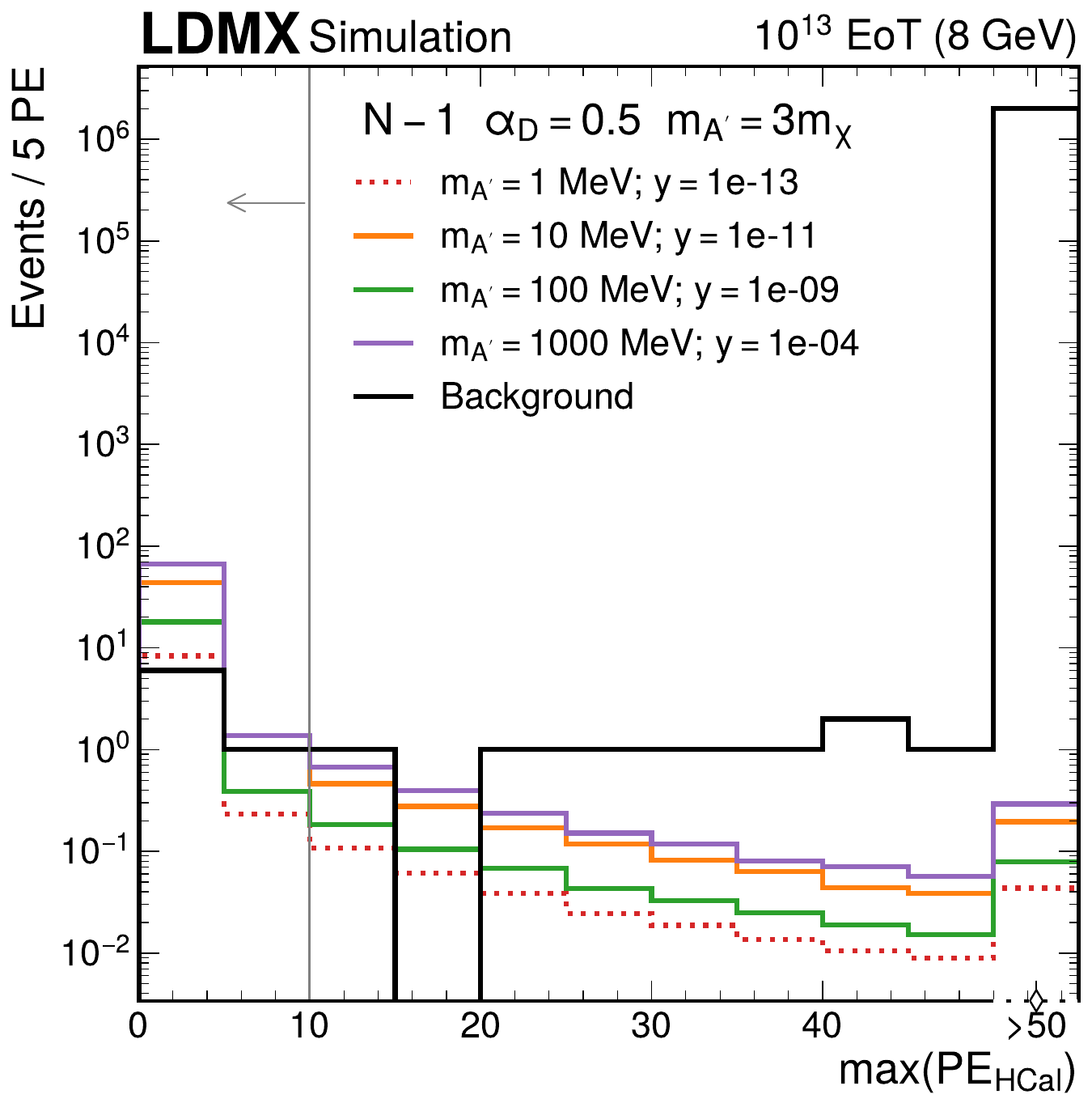} &
        \includegraphics[width=0.48\textwidth]{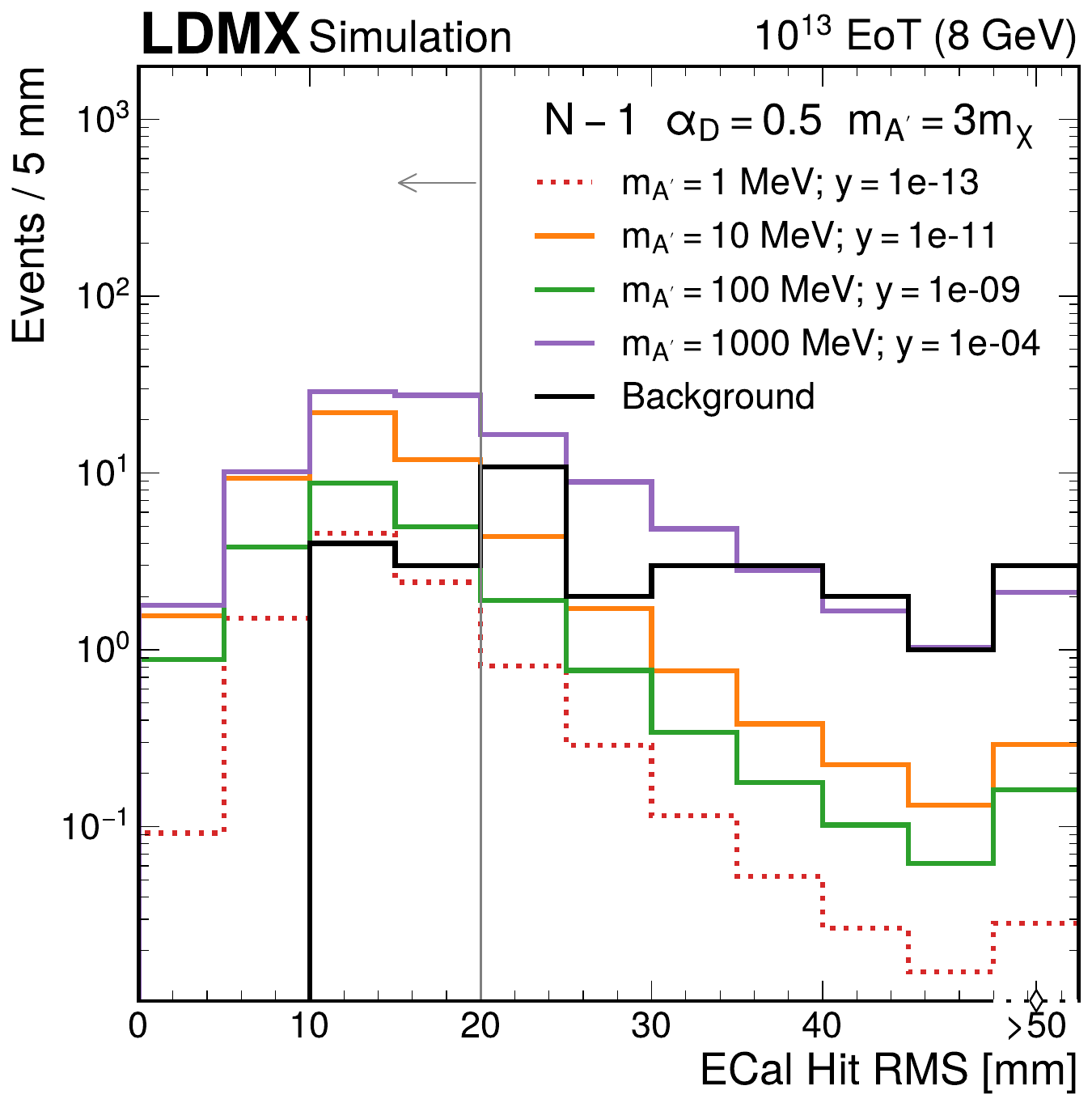} \\
    \end{tabular}
  \end{center}
    \caption{Variables used in signal selection for the \ac{eat} analysis channel.
    Each figure is an ``N-1'' plot where all other cuts are applied except for the variable in question.
    The signal samples are shown with a choice of effective interaction strength 
    $y = \alpha_D \epsilon^2 (m_\chi/m_{A'})^4$ to which this analysis is sensitive.
    For empty bins, the uncertainty shown is derived from the Poisson uncertainty based
    on nearby bins.
    The gray line shows the selection cut.
    The background event counts correspond to the enriched nuclear and di-muon samples.
    The top (bottom) row shows the \fourgev (\eightgev) beam.
    }
    \label{fig:variables}
\end{figure}

The variables used for selection are shown in Figure~\ref{fig:variables}.
In this figure, the distributions are shown after all other selections are applied
(including the trigger selection and the selection on the total \ecal energy including all layers).
The gray line indicates the selection on the observable in question, and
the background sample is composed of the enriched nuclear and di-muon samples.

\begin{figure}[htb]
  \begin{center}
    \begin{tabular}{cc}
       \includegraphics[width=0.48\textwidth]{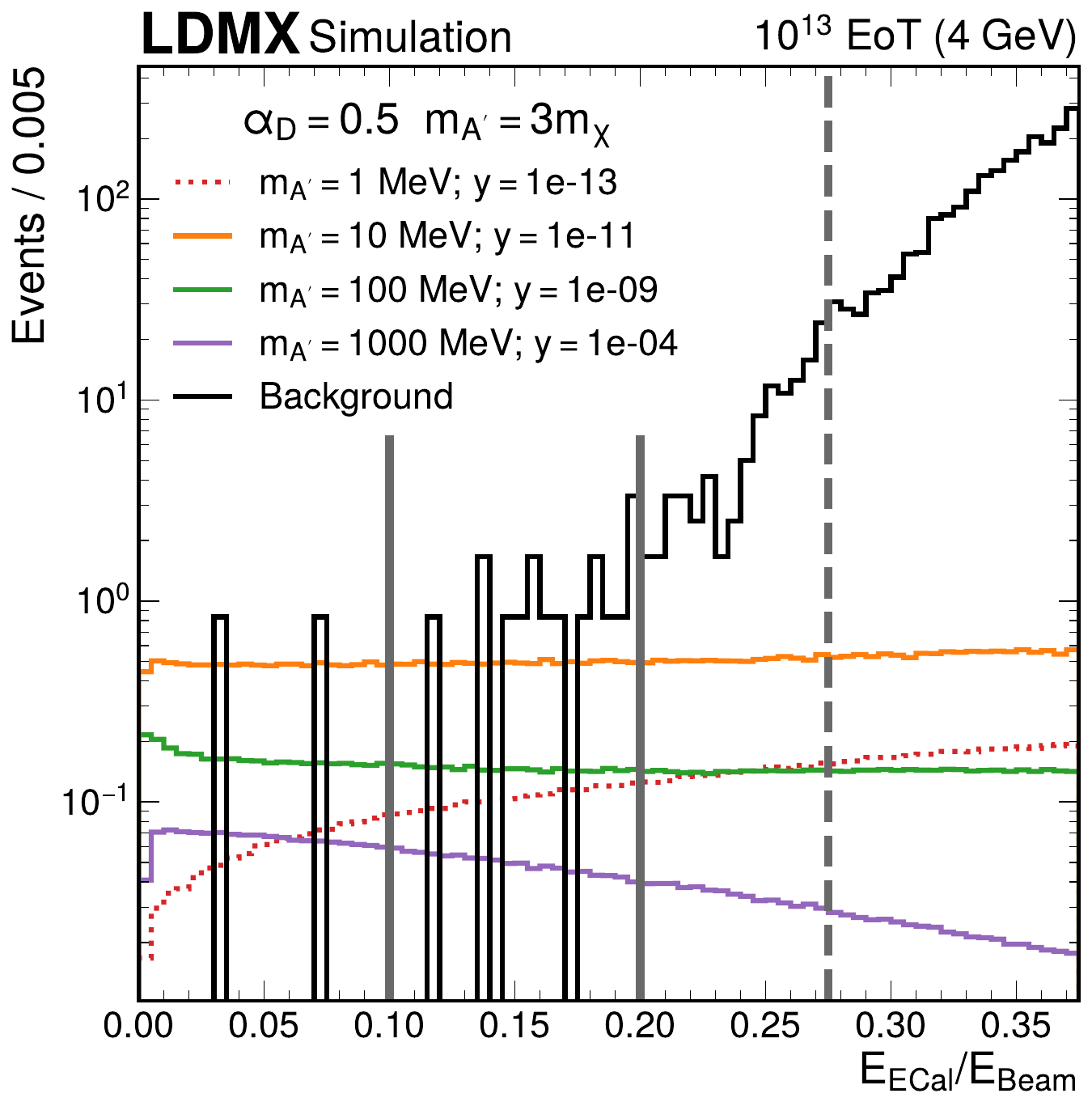}
       &
       \includegraphics[width=0.48\textwidth]{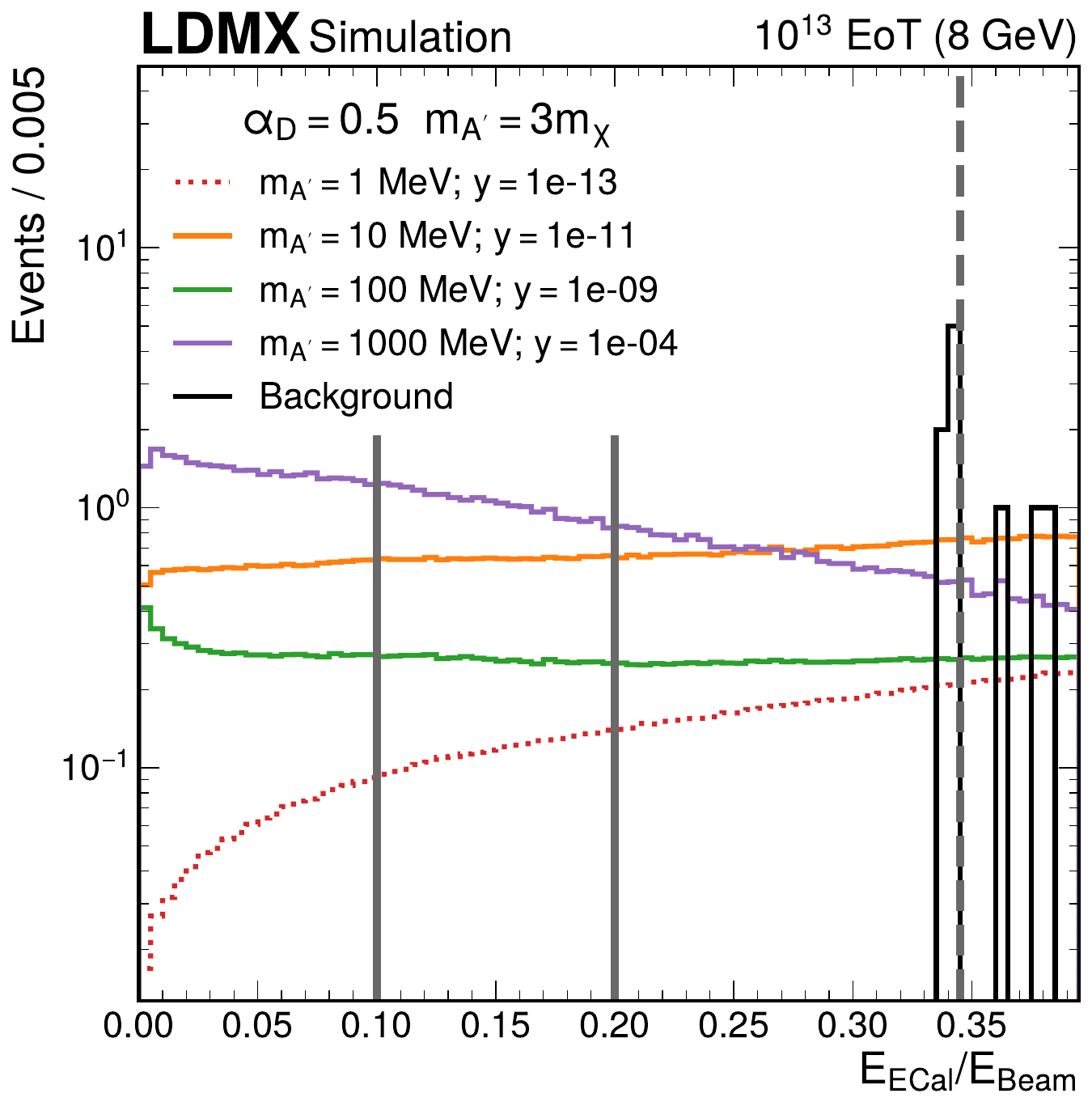}
    \end{tabular}
  \end{center}
    \caption{The total reconstructed energy in all layers of the ECal ($E_\text{ECal}$)
    as a fraction of the beam energy ($E_\text{Beam}$) for all samples that
    pass the selection criteria except the selection on ECal energy.
    The \fourgev beam is shown on the left and the \eightgev beam is shown on the right.
    The gray lines mark the edges of the analysis bins used to
    estimate the expected exclusion limit and the dashed line is the upper limit
    on the ECal energy which also serves as the upper limit of an analysis bin.
    }
    \label{fig:full-selection}
\end{figure}

The final selection is tabulated in Table~\ref{tab:cutflow}.
While not using the full power of the \ac{ldmx} detector design, these simple cuts still reject enough background while keeping signal efficiencies high.

\begin{table}[htb]
    \resizebox{\textwidth}{!}{\input{cutflow-with-mock-tracker-req}}
    \caption{
      Cut flow analysis comparing background and various signal hypotheses
      for the simple cuts used in this analysis.
      The event yield for the background sample is calculated using the event weights
      and represent the number of events out of $10^{13}$ \ac{eot} equivalent.
      The signal efficiency is relative to the simulation sample.
      The efficiency and event yield values on a given row are reported \emph{after}
      the analysis stage of that row.
      The top table is the cut flow for the \fourgev beam,
      and the bottom table is for the \eightgev beam.}
    \label{tab:cutflow}
\end{table}

Figure~\ref{fig:full-selection} shows the signal and background
total \ecal energy distributions as a fraction of the beam energy after all selections are applied.
With this selection of events, we can use the differing shapes between signal (close to uniform) and
background (sharply falling) distributions in order to further improve the signal sensitivity.
We fit the cumulative distribution of the background sample with a simple exponential function
of the missing energy fraction and then use this fit to estimate the background yield within
each of the three analysis bins.
This fit utilizes the region above the final analysis threshold (dashed line in Figure~\ref{fig:full-selection})
and below the trigger threshold to help constrain it.
Table~\ref{tab:bkgd-analysis-bins} shows the results of this calculation.
Moreover, this technique is well suited for first-contact with real data since the region currently
constraining the fit would act as a control region enabling a data-driven background prediction.
This fit is then used to give a background estimate and statistical uncertainty on this estimate
in the three final analysis bins (edges shown in gray in Figure \ref{fig:full-selection}).

Correlated systematic uncertainties on this background prediction were also estimated.
Systematic uncertainties arising from 10\% uncorrelated \ecal mis-calibrations were estimated at 5\% for all three final bins.
Uncertainty in the background rate from 10 \ac{pe} fluctuations of the maximum \ac{pe} deposited in the \hcal was also determined, resulting in no observable uncertainty in the lowest reconstructed \ecal energy analysis bin and 60\% (70\%) in the middle (high) analysis bins.
For the tracking requirements, while some misalignment may be expected for early analyses,
this analysis does not see a strong dependence on the tracking requirement and so it will be set
such that 90\% of tracks connected to full-energy clusters pass it.
Based on simulation studies, the analysis will require a tracker performance reliable-enough to reject tracks below half the beam energy, which can be achieved with minimal alignment, so no systematic uncertainty is considered for the tracker.

\begin{table}[htb]
  \begin{center}
    \input{bkgd-analysis-bins}
  \end{center}
  \caption{Background prediction and its uncertainty on the number of events within the analysis bins.}
  \label{tab:bkgd-analysis-bins}
\end{table}

%% file: cutflow-with-mock-tracker-req.tex
\begin{tabular}{|r|c||c|c|c|c|}
    \hline
    \multirow{2}{*}{Analysis Stage for \fourgev Beam} & 
      Background & 
      \multicolumn{4}{c|}{Signal Efficiency (\%)} 
      \\ \cline{3-6} 
    & Event Yield & $1$~MeV & $10$~MeV & $100$~MeV & $1$~GeV \\ \hline
    \ecal Trigger ($E_{20} < 1.5$~GeV) &
      \num{5.11e+07} & 58 & 67 & 71 & 83 \\
    Tracker Requirement ($E_e > 3.5$~GeV) &
      \num{4.60e+07} & 52 & 60 & 64 & 75 \\
    \ecal Energy ($E_{\mathrm{\ecal}} < 1.1$~GeV) & 
      \num{1.95e+06} & 32 & 43 & 48 & 65 \\
    $\max(\text{PE}_{\text{\hcal}}) < 10$ &
    \num{1.15e+03} & 31 & 42 & 47 & 62 \\
    RMS Event Size $< 20\;\mathrm{mm}$ &
      126 & 25 & 33 & 37 & 30 \\
    \hline
    \hline
    \multirow{2}{*}{Analysis Stage for \eightgev Beam} & 
      Background & 
      \multicolumn{4}{c|}{Signal Efficiency (\%)} 
      \\ \cline{3-6} 
    & Event Yield & $1$~MeV & $10$~MeV & $100$~MeV & $1$~GeV \\ \hline
    \ecal Trigger ($E_{20} < 3.16$~GeV) &
      \num{6.78e+07} & 66 & 74 & 79 & 89 \\
    Tracker Requirement ($E_e > 7$~GeV) &
      \num{6.10e+07} & 59 & 67 & 71 & 80 \\
    \ecal Energy ($E_{\text{\ecal}} < 2.76$~GeV) &
      \num{6.88e+06} & 47 & 57 & 62 & 76 \\
    $\max(\text{PE}_{\text{\hcal}}) < 10$ &
      31.8 & 45 & 55 & 60 & 73 \\
    RMS Event Size $< 20\;\mathrm{mm}$ &
      7 & 39 & 47 & 50 & 47
    \\ \hline
\end{tabular}

%% file: bkgd-analysis-bins.tex
\begin{tabular}{|c|c|c|c|}
  \hline
  Beam Energy & {Bin 1} & {Bin 2} & {Bin 3} \\ \hline
  \fourgev & $0.87 \pm 0.09$ & $15.6 \pm 0.9$ & $127.6 \pm 3.4$ \\ \hline
  \eightgev & $0.08^{+0.25}_{-0.08}$ & $0.4^{+0.8}_{-0.4}$ & $4.7 \pm 1.2$ \\ \hline
\end{tabular}

%% file: results.tex
\begin{figure}[htb]
    \begin{center}
      \includegraphics[width=0.48\textwidth]{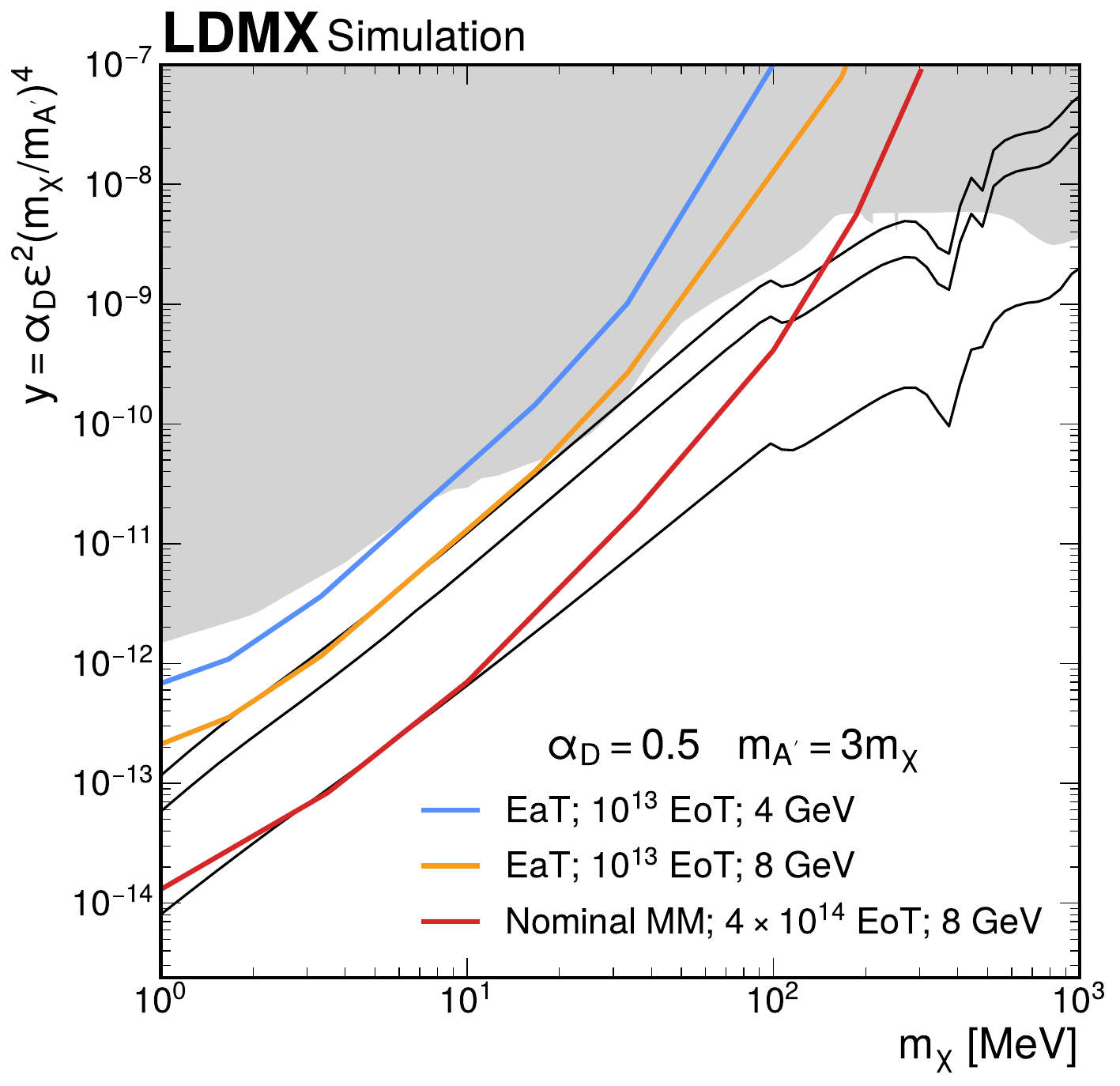}
    \end{center}
    \caption{
      The 95\% confidence level sensitivity of the \ac{eat} analysis for the \fourgev (blue) and \eightgev (orange)
      beam cases compared to other experiments led by NA64 \cite{na64_results_2019}, \babar \cite{babar_2017},
      and COHERENT \cite{coherent_2023} (gray);
      Scalar, Majorana, and Pseudo-Dirac theory expectations (black, top-to-bottom);
      and the \ac{ldmx} \ac{mm} analysis sensitivity (red).
    }
    \label{fig:reach}
\end{figure}

The expected reach of the \ac{eat} analysis in the $y-m_{\chi}$ plane is shown in Figure \ref{fig:reach}
which used the \textsc{Combine}\cite{cms-combine} statistical framework to perform a 3-bin
expected exclusion estimate using the CLs criterion \cite{cls_1999,cls_2002} with the modified
profiled likelihood ratio \cite{likelihood_ratio_2011} as test statistic and using a log-normal
model for nuisance parameters affecting yields.
The signal yield limit estimated by \textsc{Combine} is converted to a limit on $y$ using 
an estimate of the expected production rate of dark photons from the simulation described above.
This rate estimate is taken with a 20\% systematic uncertainty representing uncertainty in the effective target material composition.
The leading experimental systematic uncertainty is the HCal veto efficiency, which has a 10\% impact on the expected sensitivity.

The \ac{eat} lines reflect the results from this study for the \eightgev (orange) and \fourgev (blue) beams.
The red line is the \ac{ldmx} \ac{mm} analysis sensitivity \cite{pnpaper,pn8gev} for a total exposure of \num{4e14} electrons-on-target.
The \ac{eat} analysis is expected to achieve world-leading sensitivity in the early stages of \ac{ldmx} operation, and to extend that sensitivity as the data sample increases.  The \ac{mm} search provides additional measurement possibilities to confirm an observation of \ac{db} such as the transverse-momentum distribution of the recoil electron, which can not be precisely measured in the \ac{eat} analysis channel.

%% file: Conclusion.tex
The Light Dark Matter eXperiment will use a fixed-target,
missing-momentum approach to perform a search across a range of mixing strengths
for thermal-relic \ac{dm} with a mass of \qty{1}{\MeV} to \qty{1}{\GeV}.
This paper has described the use of the \ecal as
a secondary active target, which allows world-leading sensitivity in
the early running stages of \ac{ldmx} and provides a second analysis
channel with different systematic uncertainties and background
characteristics compared with the primary missing-momentum search
channel.  A detailed evaluation of this \ac{eat} analysis was
described, employing a limited set of selection variables appropriate
for use during early data-taking and evaluating the impact of a range
of systematic uncertainties on the analysis result.  These variables
are sufficient to suppress the known background processes while
maintaining substantial signal efficiency.  The final background
prediction and statistical analysis follows a structure that is
applicable to beam data, preparing the collaboration to conduct dark
matter physics searches with a novel apparatus with only a few weeks
of beam time.

As the beam sample size collected by \ac{ldmx} grows, the \ac{eat} analysis
channel is anticipated to maintain comparable exclusion sensitivity to
the \ac{mm} analysis by having access to a higher signal rate and
through the staged introduction and validation of more advanced
variables to suppress the backgrounds which will become relevant for
larger detector exposures.  The \ac{eat} analysis is therefore
expected to complement the \ac{mm} analysis and extend the reach
of \ac{ldmx} data.